# Political and Economic Patterns in COVID-19 News: From Lockdown to Vaccination

**ABDUL SITTAR[1], DANIELA MAJOR[2], CAIO MELLO[3], DUNJA MLADENIIĆ[4], and MARKO GROBELNIK[5]**

[1,4,5]Department for Artificial Intelligence, Jozef Stefan Institute, Jamova cesta 39 1000, Slovenia (abdul.sittar@ijs.si, dunja.mladenic@ijs.si, marko.grobelnik@ijs.si)

[2,3]School of Advanced Study, University of London, Senate House, Malet St, London WC1E 7HU, United Kingdom (daniela.major@sas.ac.uk, caio.mello@sas.ac.uk)

Corresponding author: Abdul Sittar (abdul.sittar@ijs.si).

The research described in this paper was supported by the Slovenian research agency under the project J2-1736 Causalify and by the European Union's Horizon 2020 research and innovation program under the Marie Skłodowska-Curie grant agreement No 812997.

**ABSTRACT** The purpose of this study is to analyse COVID-19 related news published across different geographical places, in order to gain insights in reporting differences. The COVID-19 pandemic had a major outbreak in January 2020 and was followed by different preventive measures, lockdown, and finally by the process of vaccination. To date, more comprehensive analysis of news related to COVID-19 pandemic are missing, especially those which explain what aspects of this pandemic are being reported by newspapers inserted in different economies and belonging to different political alignments. Since LDA is often less coherent when there are news articles published across the world about an event and you look answers for specific queries. It is because of having semantically different content. To address this challenge, we performed pooling of news articles based on information retrieval using TF-IDF score in a data processing step and topic modeling using LDA with combination of 1 to 6 ngrams. We used VADER sentiment analyzer to analyze the differences in sentiments in news articles reported across different geographical places. The novelty of this study is to look at how COVID-19 pandemic was reported by the media, providing a comparison among countries in different political and economic contexts. Our findings suggest that the news reporting by newspapers with different political alignment support the reported content. Also, economic issues reported by newspapers depend on economy of the place where a newspaper resides.

## I. INTRODUCTION

We can say that news coverage directly catalogs the occurrence of specific events and indicates the local as well as global opinions of stakeholders. As the epidemic took over the world, news coverage became a significant source of information that allowed populations to adapt their behaviours and fight off the disease. Because of this, COVID-19 has been extensively reported by Global and local media [1]. Analysing the multi-facet and spatio-temporal aspects of the news coverage of the pandemic is essential for a clear understanding of this event, as news contributes to the way people understand the world. It is a 'place of reference' [2] where people 'go' in search of stability. In order to understand the evolution of the coverage of this pandemic, we propose a methodology in which we employ variants of popular techniques of Natural Language Processing (NLP) such as Topic Modeling (TM) and Sentiment Analysis (SA). Topic models are designed to analyze a corpus of text and extract the latent themes or topics. The results are more coherent if the text is semantically similar. Therefore, the motivation behind our proposed method is to pool the news articles based on user queries to extract the most relevant latent themes without modifying the basic structure of LDA. Previous studies used pooling based on other parameters. Mehrotra et al. applied pooling based on hashtags on twitter datasets [3], while [4] proposed a scheme for pooling tweets into longer documents based on conversations. However, the problem of pooling based on user queries is not explored for news articles. To fulfil this gap, the present study aims to see the coherence score differences with and without pooling based on user queries. SA is used to check positive or negative sentiments within a text. Among famous sentiment analyzers (Ratio, TextBlob, NRC Emotion Lexicon, VADER), we use VADER to extract sentiments. The motivation behind finding the sentiments is to provide information on how the media approached each phase of the pandemic. The identification of sentiments discursively expressed in the narration of this event offer insights on how different outlets interpreted the different dimensions of the pandemic such as the lockdown, the protective measures, the research to develop vaccines and their distribution over time.

Numerous studies have investigated the impact of COVID-19 in different countries. [5] investigates how different discussions evolved over time and the spatial analysis of tweets. [6] addresses the diffusion of information about the COVID-19 using a large amount of data from popular social media networks. [7] performs SA in the early stages of the COVID-19. These studies, however, do not conduct experiments on large timespans that may provide a better overview of the pandemic. Secondly, most studies employ Twitter content, and when news are used they are either limited to one or two newspapers or belong to the same country.

This study identifies how the discussions evolved over time in top newspapers belonging to three different continents (Europe, Asia, and North America) and nine different countries (UK, India, Ireland, Canada, U.S., Japan, Indonesia, Turkey, and Pakistan). It uses spatio-temporal TM and SA. TM will be used to determine the topics of discussion especially pertaining to different economic and political perspectives, while SA will be used to determine the change in sentiments over time. Identifying these topics and emotions could help newspapers to improve how they communicate information and provide data which would enable governmental bodies review their communication strategies.

The remainder of this paper is structured as follows: Section II provides related work on spatio-temporal analysis, SA and data description about COVID-19. After elaborating on the research methodology in Section III, the results are described in Section IV. Section V provides a brief discussion about findings and the corresponding results. Finally, Section VI presents the conclusion and some ideas for future work.

### A. MOTIVATION

The motivation behind our work are stemmed from the following facts:

- With a rapidly growing number of events with significant international impact, understanding the differences in news reporting has increased importance for researchers and professionals in many disciplines, including digital humanities, media studies, and journalism. The most prominent recent examples of such events include the migration crisis in Europe, COVID-19 outbreak, and Brexit.
- There are many factors that influence the news selection, reporting, and spreading such as cultural, political, economic, geographical, and linguistic. Analyzing these factors in news spreading related to different international events is open research area. Since COVID-19 news has many conspiracies and fake information attached with it, and it has major effects on different economies, we take into account only two factors political and economic.
- Nowadays, social scientists and psychologist are interested to know how the COVID-19 pandemic has influenced the everyday life of people in the world. It has raised the issues including unemployment, stress and depression due to the lockdown, and inflation and so many other issues. In general, newspapers report overall situation of a specific area. Analyzing the sentiments in news articles can help to see the changes in sentiments over time and to make comparison of sentiments across different countries.

### B. CONTRIBUTIONS

The following are the contributions of this study:

- We propose and evaluate the enhanced Topic Modelling approach that uses LDA with combination of 1-6 grams and articles' pooling based on queries to improve the quality of topics without modifying the basic structure of LDA.
- We propose a methodology to understand political and economic differences in news reporting using TM.
- We study the comparison of sentiments between newspapers across different geographical areas.

### C. RESEARCH QUESTIONS

This paper is the first of its kind to analyse the news related to COVID-19 based on political alignment and the economic situation of different countries from January 1st, 2020 to May 31, 2021. Our hypothesis states that the topics present in news related to the COVID-19 pandemic vary according to the publisher's political alignments and the economic situation of a specific area. Moreover, the topics that have strong relation with each other will have similar trends over time. We used TM, and SA to answer the following research questions:
**Q1:** How political and economic issues propagated over time during the pandemic across different socio-political and economic contexts?
**Q2:** How different discussions evolved during different stages of the COVID-19 epidemic?
**Q3:** What are the patterns of emotional states during different stages of the pandemic across different countries?

In our first research question, we refer "political issues propagated over time" as political issues that have spread/reported over time. Similarly, "economic issue propagated over time" refers to the economic issues that have spread/reported over time. In our second research question, "discussions" refers to different topics that evolved during different stages of the COVID-19 epidemic. In our third research question, "emotions" refers to sentiments over time.

## II. RELATED WORK

COVID-19 is a broad topic and has enormous number of research dimensions. As this study focuses on analyzing the spatio-temporal political and economic patterns, and sentiment analysis, we review six different types of related works: data for COVID-19 analysis, spatio-temporal analysis, COVID-19 analysis using social media, characteristics of the newspapers, sentiment analysis, and topic modeling.

### A. DATA FOR COVID-19 ANALYSIS

Analytical studies regarding COVID-19 have conducted research to portray its emerging effects in different fields using popular analytical methods and data from different sources such as Nigeria Centre for Disease Control (NCDC) [8], Facebook [9], News from the New York Times (United States of America) and Global Times (China) [10], Twitter [11], [12], News from China National Knowledge Infrastructure (CNKI) database [13]. The timeline for all this data coverage is only a few months (two, three, or the first few months of the pandemic). [14] has studied general issues reported in news media. To the best of our knowledge there is a lack of studies that cover the complete phase of the pandemic from lockdown to the vaccination (from January 2020 to May 2021) by using news and that find the issues focusing on different political and economic contexts.

### B. SPATIO-TEMPORAL ANALYSIS

Spatio-temporal analysis is used to uncover the relations between locations over time [15]. [16] identify spatial effects and spatio-temporal patterns of the outbreak of COVID-19 in different regions of Italy. [17] identifies seasonality in disease in Spain due to variations in temperature, humidity, and hours of sunshine. [18]. [19] study the spatio-temporal propagation of the first wave of the COVID-19 virus in China and compare it to the other global locations in terms of distance, population size, and human mobility and their scaling relations. To our knowledge, there is no study which focus on top read newspapers. And these newspapers belong to nine different countries which belong to three continents.

### C. COVID-19 ANALYSIS USING SOCIAL MEDIA

Social media is one of the major sources to understand the societal and crowd response nowadays [12]. One of the reasons is its widespread growth. However, the consumption of news articles through news publishers is associated positively with higher trust whereas the information related to the social media is linked with lower trust [20]. Analyzing and linking emerging events to relevant social issues is an increasingly important task. Most of the focus has been on social networks when studying spatio-temporal effects of the COVID-19 Pandemic. News is also one of the most important sources containing detailed information [21]. There is a fundamental problem attached to news articles which relates to the unstructured and noisy nature of data. . Several studies employ news but are limited to one or two countries. [22] measure depressiveness, and informativeness for various states in the US. [10] take news from the New York Times (United States of America) and Global Times (China) to study the way the news used for political and ideological purposes. To our knowledge, there is space for research studies which conduct COVID-19 spatio-temporal analysis by utilizing news published around the world (including countries from different continents).

### D. CHARACTERISTICS OF THE NEWSPAPERS

International news led us to investigate the reasons why specific event-centric news either spread or do not spread to certain geographic area. Based on some factors, media target specific foreign and regional events. For example, spreading of news related to specific events may tilt toward developed countries such as United Kingdom, U.S.A, or Russia. Furthermore, it may be due to geographical juxtaposition (latitude, longitude) of countries [23]. There is great deal of negotiation between political actors and journalists in news production to enhance their influence on news coverage. Fake news also produced based on many factors and surrounded by a dominant element that is political effect [24]. Thus, political alignment of news publishers also impact the coverage of different events. One of the determinants for news coverage is economic conditions that also impact information selection, analysis, and propagation [25]. Generally, news reporting about different events (elections etc.) is inclined towards certain characteristics of newspapers. As there is a inclination to back underground and indirectly, the research interest in the reporting characteristics of each newspaper has begun [26]. Filla investigates the political participation by the local news outlets in elections and find the relationship between the political participation and availability of local news outlets [27]. News agencies tend to follow the national context in which journalists operate. One of the related examples is the SARS epidemic study which found that cross-national contextual values such as political and economic situations impact the news selection [28]. A great amount of work regarding fake news dwells on different strategies, while few studies considered political alignment to have a compelling effect on news spreading [29]. With a rapidly growing number of events with significant international impact, understanding the news reporting has increased importance for researchers and professionals in many disciplines, including digital humanities, media studies, and journalism. The most prominent recent examples of such events include the migration crisis in Europe, COVID-19 outbreak, and Brexit. Although the previous work involves relationship between outlets and political activities or public interest [30], our work focused directly on studying insights in reporting differences in different political and economic contexts.

### E. SENTIMENT ANALYSIS (SA)

SA is used to check different levels of positive or negative opinions within a text. It is useful to determine the emotional state expressed in news articles in response to the outbreak. [5] performed SA on tweets belonging to different countries and generated time-series plots to see whether the spikes in positive or negative sentiment can be associated with certain events. Another study develop a Recurrent Neural Network (RNN) for predicting emotions using tweets and compare the model with TextBlob [31]. Some studies find country wise sentiment analysis using R [32], NRC Emotion Lexicon [33]. [12] highlight the concerns of Gulf countries' people to lessen the vaccine hesitancy. It uses three different methods (Ratio, TextBlob, and VADER) to extract the sentiments. Working with SA implies a range of challenges to obtain accurate results such as the analysis of negation, sarcasm and ambiguity [34]. These limitations were considered when making use of this algorithm. The technique is used as a exploratory tool to produce insights of the data rather than a conclusive method.

[35] use a method that connects the polarity scores to emotional states with the use of Emotional Guidance Scales. This scale from -1 to 1 contains 11 emotions where each emotions change with increments of 0.2. -1 denotes the most depressed and fear feeling and +1 denotes emotion of being happy and joyful. This brings up the interesting problem of spatio-temporal SA from news articles. The objective is to discover sentiment on news articles on the COVID crisis at country level over temporal intervals of a month using a large collection of news articles from a period starting from January 2020 to May 2021.

### F. TOPIC MODELING (TM)

Several studies used TM to determine the topics of discussion about COVID-19. The intention was to extract popular topics and understand evolution of different discussions [36], [37], [38]. LDA is used to infer topics from the collection of text-document. Some techniques used only frequent words whereas some use pooling to generate relevant topics and maintain coherence between topics [5]. To pool the relevant documents several mechanisms have been used. Unlike simple static TM in this modeling it is assumed that the data is partitioned on a time bases (e.g. hourly, daily, monthly, or yearly). In fact, the order and arrangement of documents reflects the evolving set of topics [39]. These pooling techniques are famous for social media where set of documents/tweets are partitioned based on hashtags, and authors, etc. [3]. Carmela combines peak detection and clustering techniques to identify topics. Space-time features are extracted from tweets and modeled as time series. Using these time series peaks are identified and clustered based on co-occurrence in the tweets [18]. Innovative approaches have been proposed to detect spatio-temporal topics. The problem of topic identification over spatio-temporal analysis is also considered as problem of stream clustering. Tweet-based clustering method used the content, structural and temporal information, Hashtag-Based clustering focused on hierarchical spatio-temporal techniques which explore different event with different time granularity [40]. Information retrieval is understood as an automatic process that respond to a user and returning a list of documents that should be relevant to the user query [41]. In this process, terms (which appear in the queries and set of documents) are ranked using some weighting techniques. One of the popular methods is TF-IDF weight. This is a statistical measure used to evaluate how important a word is in a document in a dataset [42]. Data provided by social media can be pooled based on meta data (e.g. hash tags in Twitter) but for news articles this type of pooling does not exist. Therefore, we pool the news articles based on user queries. LDA is used to discover underlying topics directly from the raw text features in the documents. When it is applied directly on raw texts then result in topics are uninformative and hard to interpret [4]. Pooling of raw text into groups appeared appealing with vast improvements in results [43]. However, these methods only applied on data which has structured form of schema such as Social Networks (SN). The information in SN has attractive options such as share or retweet and information makes cascading structure, which means one can easily group the information based on the users, hashtags, timestamp and location. Contrary to SN, there is no such structure exist in the case of news articles. Therefore, to overcome this issue [14] uses Top2Vec algorithm and Doc2Vec to place news articles close to other similar news articles. The limitation of this study is that it requires to input all the news articles at the same time while finding the topics. Although the methods used are efficient, the results would change if news articles grouped based on user queries. [44] used topic modeling to depict the overall picture of Taiwan's pandemic where they divide the data into different stages based on temporal information. They divide the data into four stages based on the development of COVID-19. They have also used basic LDA technique and the quality of topics is compromised as the news articles are not filtered based on queries or general themes. Our study provides the basis for improvement of the results if we group the news articles based on queries.

## III. METHODS

This study adopts an approach that integrates computational as well as qualitative techniques to explain perceptions and attitudes towards COVID-19. The present research focuses on spatio-temporal analysis of different discussions: news events and governmental decisions, cures and prevention measures, political and economic discourses, and emotions. Fig. 1 shows the experimental methodology adopted in this work. Several steps were performed sequentially in the workflow, including data collection, data preparation, filtering news articles, TM, topic analysis, analysis of political and economical issues, and SA. This methodology can be adopted on news of different type of events to understand the effects of different political and economical context's on news reporting. The following sections present the data collection and preprocessing, TM, topic visualization, and SA in detail.

### A. DATA COLLECTION AND PREPROCESSING

The collected dataset of news articles is based on ten newspapers [1]. These newspapers were selected based on the preconditions:

- At least a few news articles must be published by a newspaper for each month (January 2020 to May 2021),
- Newspapers should belong to different political alignments,
- The newspapers should belong to countries from different economic backgrounds.

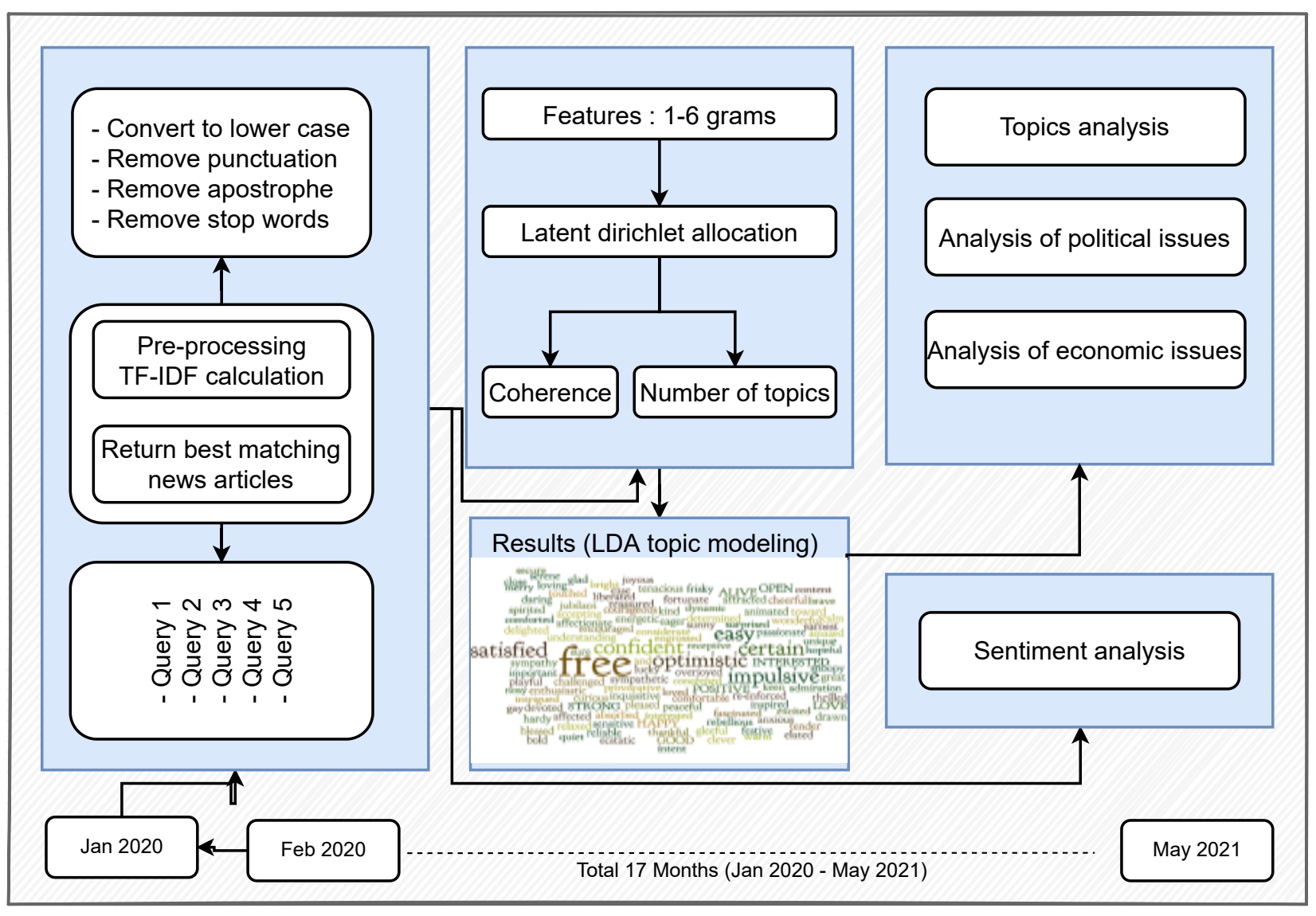

**FIGURE 1. Workflow to identify patterns of different political and economic issues as well as SA. Data preprocessing for five different queries for each month (sequentially) is the initial step. TM using combination of 1-6 grams is the second step. The last step is to identify the most frequent political and economic issues. In case of SA, the body text of news article temporally pass to the sentiment analyzer.**

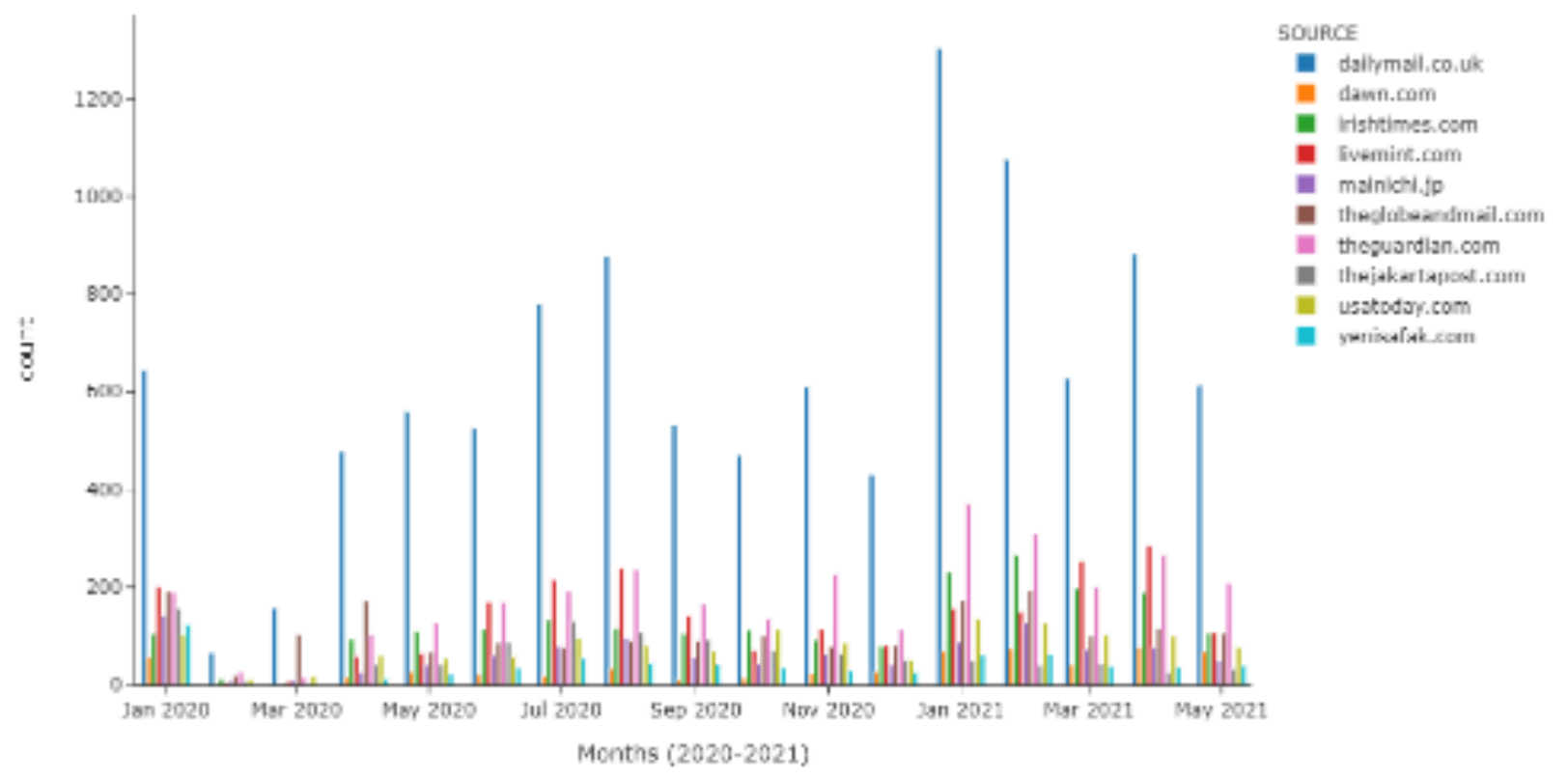

**FIGURE 2. Total number of news articles published in each month (01/2020- 05/2021) by ten newspapers. Main purpose of the line graph is to show the variations in collected number of news articles reported by top ten newspapers during the set timeline.**

Following these preconditions we constructed a dataset which consists of 24,000 news articles published by ten newspapers. Figure 2 shows the number of articles published by each newspaper in each month. Figure 3 shows coverage of news articles per country, continent and political alignments. The data was collected by using Event Registry platform. This platform identifies events by collecting related articles written in different languages from tens of thousands of news sources [45]. Using Event Registry APIs [2], we fetched news articles in English published by the selected newspapers. We use the following keywords to search news articles: COVID-19, Coronavirus, COVID pandemic, and COVID Outbreak. Each news article consists of a few attributes: title, body text, name of the news publisher, time of publishing. We estimate the political alignment of each newspaper using Wikipedia info-box as already done in [15]. We perform a number of preprocessing steps: conversion to lower case, removal of punctuation marks and removal of stop-words. It is important to mention that there were other newspapers (see the link [3]) that were unable to follow the preconditions. Some newspapers stood in same position in political spectrum. Some newspapers belonged to countries with same economic backgrounds. For some of the newspapers, we were unable to find news articles from Event Registry platform for few of the months in the set time period (from January 2020 to May 2021). Therefore the newspapers reduced and as a result the news articles also reduced to few thousand (24,000) news articles. Another significant information that we can see in Fig. 3 is the varying number of news articles by all newspapers. For example, in case of UK based newspaper Daily Mail, we see there is large collection of news articles in each month, whereas for U.S.A, there is only one newspaper with approximately five percent of news articles. This is because of our preconditions. If we remove this potential bias then the main goal of this methodology would not be achieved that is understanding news reporting across different political and economic contexts.

### B. TOPIC MODELING (TM)

We identify main topics related to COVID-19 and construct five queries to pool news articles for each month in the period between January 2020 and May 2021. These five queries are: 1) Lab leak theory, 2) Efficiency of vaccines, 3) Lockdown policies and efficiency, 4) Seriousness of Coronavirus, and 5) Can masks protect against COVID-19.

Previous studies conducted research on most critical topics related to COVID-19 pandemic ( [46]–[50]). We construct queries following these famous topics. Since these are the most researched topics related to COVID-19, we select them to see the reporting differences across difference newspapers. To pool relevant news articles, we perform filtering based on each query for each month by calculating TF-IDF score of unique words for all news articles. During the filtering process, for individual newspaper and for each query on the time period of one month, we take top matching news articles (ten percent if total news articles are greater than hundred,

[1]https://www.4imn.com/top200/

[2]https://github.com/EventRegistry/event-registry-python/blob/master/eventregistry/examples/QueryArticlesExamples.pyhttps://github.co registry-python/blob/master/eventregistry/examples/QueryArticlesExamples.py

[3]https://www.4imn.com/top200/

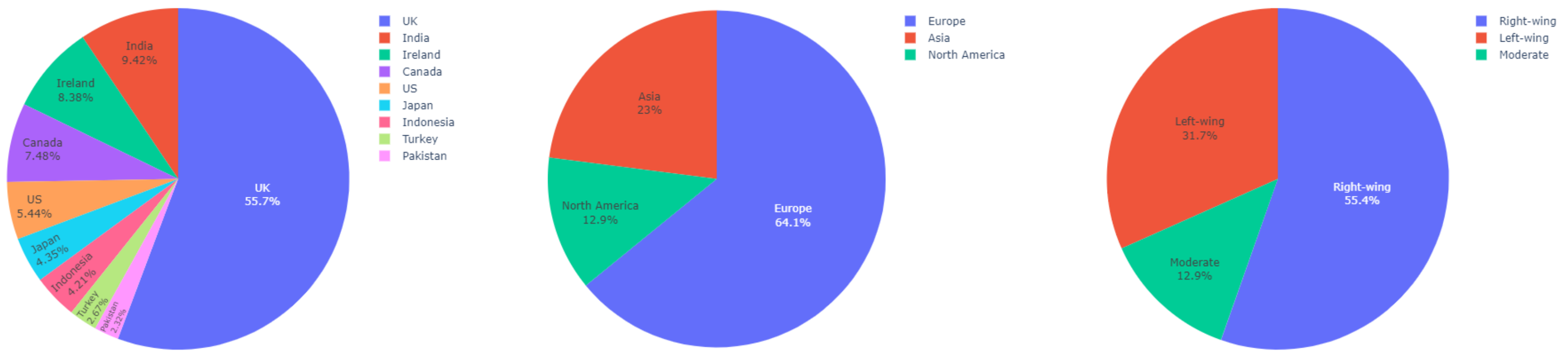

a). Percentage of news articles by nine countries.

b). Percentage of news articles by three continents.

c). Classification of newspapers with respect to political spectrum.

**FIGURE 3. Data distribution is not equal in all three cases (country wise, continent wise, and with respect to political spectrum). Statistics about newspapers w.r.t. their geographical location of headquarters and political alignments.**

ten if total news articles are less then hundred and greater than ten, all news articles if total news articles are less than 10). We perform the same preprocessing steps on each query, as we did on the news articles: conversion to lower case, removal of punctuation marks and removal of stop-words. Then for each token from the query we calculate TF-IDF score for each news article and then sort the news articles based on the sum of TF-IDF score of these tokens. Then we take into account the top news articles. For instance, we have a query "Can mask protect against corona virus?". We perform preprocessing such as conversion to lower case, removal of punctuation marks and stop-words. The final tokens for this query are four e.g., "mask", "protect", "corona", "virus". We sum the TF-IDF scores for these tokens in each of the news articles. Based on the outcome, we sort the news articles and take top news articles as relevant for the query. A method that is used for finding the abstract topic in a large collection of documents is called TM. LDA is a thematic probabilistic modeling algorithm that takes into account both words and documents while capturing the topics. We call a set of topics coherent if they support each other and cover most of the facts in a set of documents. There are many coherence measures (C_v, C_p,C_uci, C_umass, C_npmi, C_a) and the way they are calculated is different. We use only C_c measure. It is based on sliding window. It basically performs one-set segmentation of the top words and an indirect confirmation measure that finds normalized point-wise mutual information (NPMI) and cosine similarity. We compare coherence score of each individual query with query and without queries along with simple uni-gram representation for each newspaper. We see that content is more coherent with queries (see Figure 4). Also, we find an optimal number of topics for all queries and for all newspaper (see Figure 4). We use a combination of 1 to 6 ngrams when performing TM on news articles.

### C. TOPIC VISUALIZATION

We filter political and economic topics for each query manually. For each type of political alignment we put together all the filtered topics and show them in word clouds. Figure 5 shows the word clouds for different political alignments along with all the queries. Figure 6 shows the word clouds for different economic levels along with all queries. There were different number of filtered news articles for each newspaper that are used to create Figures 5, 6 and 7. More specifically, for each query, there were 615, 308, 227, 209, 355, 488, 131, 439, 112, 2125 news article filtered for Dailmail.co.uk, Dawn.com, Irishtimes.com, livemint.com, mainichi,jp, theglobeandmail.com, theguardian.com, thejakartapost.com, usatoday.com, yenisafak.com respectively. For each query the number of filtered news articles was the same within the same newspaper, but the news articles were different following the process of filtering that take queries into account. To observe the topics which had similar trend over time, we use topics that have been filtered for different political alignments and economic levels. These trends are based on frequency of these topics over time. Afterward, we generate the trend lines using the Hodrick Prescott filter [51].

### D. SENTIMENT ANALYSIS (SA)

Sentiment in news is expressed in a variety of forms, from interviewers quotes to journalists choices to use one term instead of another. This becomes clear when words such as fear are used in the headlines "Coronavirus Spreads Fear". There was a choice of using "fear", which is a word that carries negative sentiment in Vader dictionary. The sentiment of each news articles was classified using VADER Sentiment Analyzer. It is a rule-based SA tool and a lexicon which is used to express sentiments in social media [52]. In the analysis, we only take body text of news articles into account. Figures 8, 9, and 10 depict the comparison of sentiments across three continents (Asia, Europe, and North America), utilizing the news articles. The granularity of sentiment is limited to days.

## IV. RESULTS

The aim of proposing the enhanced TM approach was to improve the quality of topics without modifying the basic structure of standard LDA. For that we evaluate the model based on coherence score which shows that pooling of news article can improve the quality of topic (see Figure 4) but it also help to answer customized research questions (see

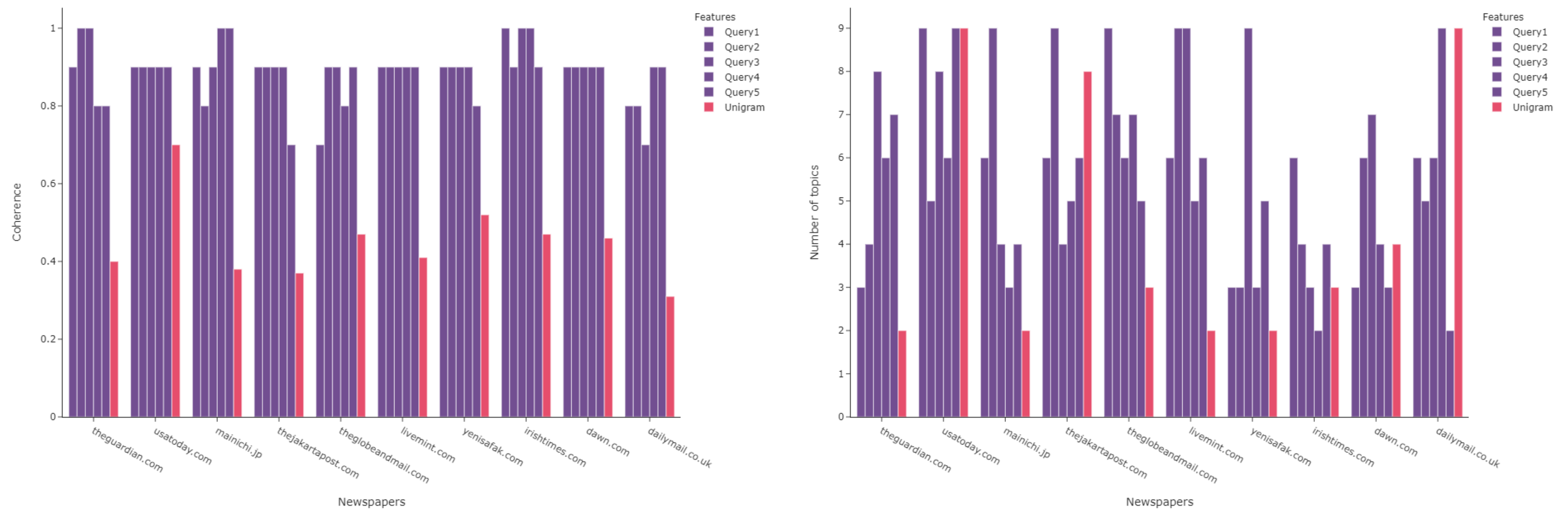

**FIGURE 4. Coherence score increases and number of topics also varies with pooling. Comparison of pooling with simple uni-grams based on C_V Coherence measure (left-hand side) and the number of topics (right hand-side). The left bar chart compares the coherence measure score between pooling with combination of 1-6 grams and simple uni-grams without pooling. Coherence score is always high in case of pooling than without pooling. The right bar chart compares the best number of topics using 1-6 grams with pooling and without pooling along with simple uni-grams. Number of topics also vary for each newspaper for all queries.**

Section IV-A, IV-B). Figure 4 shows the comparison of coherent score of LDA with pooling and without pooling. There are six bars for each newspaper. Red line shows the coherent score without pooling and other lines show coherent score with pooling for our five queries (see Section III-D). We can see that coherent score of LDA topics is always higher than 0.7 for all queries in case of pooling, whereas in case of without pooling and without queries, the coherent score of LDA topics is always lower than 0.7 for each newspaper. The second bar chart in figure 4 shows the variations in number of LDA topics with pooling and without pooling. There are six bars for each newspaper. Red line shows the number of topics for without pooling and other lines show coherent score with pooling for our five queries (see Section III-D). It shows that for seven newspapers the number of topics increased with pooling for all the queries whereas only three newspapers have higher number of topics without pooling. Overall, It shows that with pooling for each query the coherent score of LDA topics is high and number of LDA topics also increased. The aim of this spatio-temporal analysis is to answer the three research questions regarding COVID-19, that we have described in Section I-C. For the first research question related to spreading of political and economic issues, we observe the frequency of words in news articles and visualize it using word clouds containing the frequent topics across different political alignments and different levels of economies. Analysing the frequent topics across different political alignments can help us to find correlations between topics and political alignments. Therefore it might be possible to see whether political alignment is associated with particular effect on news spreading related to COVID-19 or not. Analyzing the frequent topics across different levels of economic prosperity can help us to see the correlations between certain topics and different levels of economic prosperity [4]. Therefore, it might be possible to see whether a country's economic situation has a certain effect on news spreading related to COVID-19 or not. For the second research question related to the discussions development over time, we compare trends of different political and economic terms and present as line graphs (see Figure 7). Looking at trends of words over time can help us understand how major political and economic topics evolved together. As a result we can see if there is any semantic relation between those topics or not. To answer the third research question we created a line graph for each newspaper based on sentiment score (see Figures 8, 10, 9). Sentiments over different geographical areas help to compare and rank the emotional states expressed through news articles. Moreover, looking at the overall stance of reporting by a newspaper during a specific time period requires an explanation about why at this time the reporting is positive or negative. Therefore we also identified a list of common and most frequent topics shown in Table 3.

### A. POLITICAL ISSUES:

Figure 5 shows the word clouds of main and common topics discussed in newspapers related with different political alignments. The findings have been summarized below:

**Lab leak theory:** Left wing including liberal views have been found when discussing the origin and emergence of virus, such as seafood and laboratories whereas moderate and right wing newspapers appeared to focus more on conspiracy and misinformation about virus. The conservative Islamic newspaper discuss symptoms and experiments to combat the virus.

**Efficacy of vaccines:** The newspapers representing five different political views haves shown different points of view

[4] https://datahelpdesk.worldbank.org/knowledgebase/articles/906519-world-bank-country-and-lending-groupshttps://datahelpdesk.worldbank.org/knowledge world-bank-country-and-lending-groups

on vaccine efficacy. Left wing newspapers talk about the after-effects of vaccination and raise questions about the efficacy of vaccines. Liberal newspapers forecast the rollout of vaccination and write about different vaccines along with age-group vaccination and their efficacy level. They also mention shipment of stocks and their distribution.

**Lockdown policies and efficiency:** In the case of lockdown policies, left wing newspapers are found discussing restrictions and the effects of lockdown on different fields such as tourism and industry, whereas right wing newspapers are found discuss anti-lockdown protests.

**Seriousness of COVID-19 virus:** In the case of the seriousness of COVID-19, left wing newspapers published news discussing symptom identification and the financial consequences of the pandemic, whereas liberal newspapers mention the challenges of this disease, such as worldwide confinement. They also mention the required equipment to fight off the disease such as ventilators. Moderate newspapers, on the other hand, appeared to have entirely focused on the disease symptoms, such as respiratory and heart problems and the need for oxygen. Right wing newspapers reflect topics related to the issues faced by the general public as a consequence of lockdown, including shut down places, anti-lockdown rallies and pandemic forecastings.

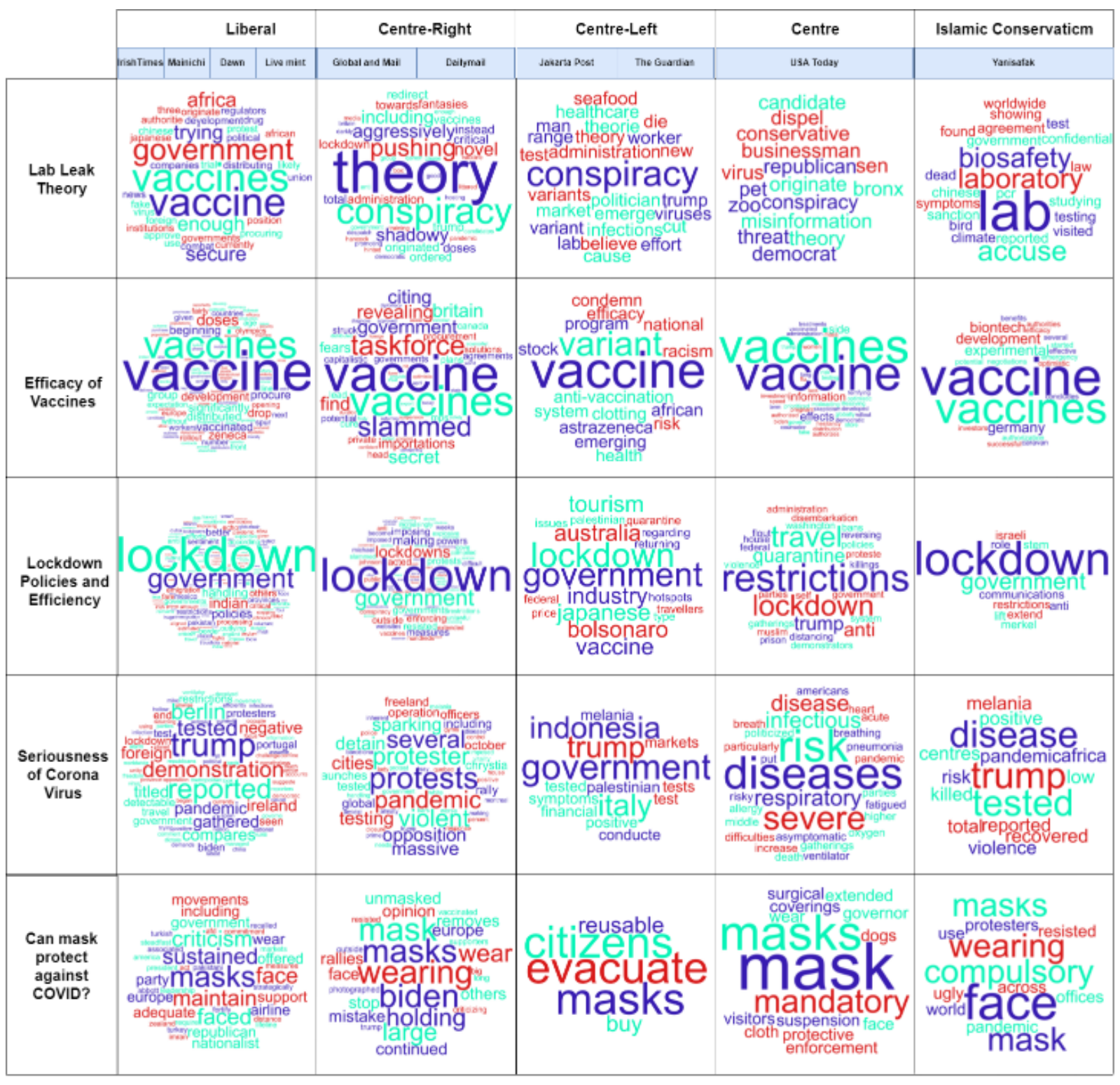

**FIGURE 5.** Word clouds showing the keywords appearing most frequently related to five queries in relation to different political alignments.

**Can mask protect against COVID?:** When it comes to mask protection, left wing newspapers discuss reusable masks. Liberal newspapers discuss mandatory face masks requirements and social distancing at airports. Moderate newspapers talk about a variety of masks, such as cloth masks. Right wing newspapers publish news about changes in policy in Europe when it comes to face masks. For example, how vaccination changes the mandatory requirements for mask wearing. Islamic conservative newspapers resist compulsory mask wearing.

### B. ECONOMIC ISSUES:

Figure 6 shows the word clouds of main and common topics discussed in newspapers belonging to different economies. Each topic has been discussed in different economies with different results. The findings have been summarized below:

**Lab leak theory:** High income countries mention the topics theory, practice, pitfalls, scheme in connection to the lab leak theory, which might indicate that they engage in discussions about conspiracy theories. As middle income countries benefit from trade and investments made by high income countries topics such as funds, trade and investors appeared with high frequency. Generally, lower income countries struggle financially so the content published and discussed in the news is more related to funding, pays, pricing, therefore topics such as farmers, banking, and pricing appeared as most frequent.

**Efficacy of vaccines:** The terms that appeared amongst low income countries are vaccine unions, cost and potential marketing. It shows that low income countries are potentially more concerned about affording the vaccine. In the case of middle income countries, we see economy, trade, vaccine roll-out and administration costs were more prevalent topics, which shows, as previously said, that middle income countries are more concerned with trade. As high income countries such as U.S and China were more involved in vaccine production, we see topics related to vaccines such as number of doses, stocking the vaccine and large task forces.

**Lockdown policies and efficiency:** We see that all countries were discussing issues that occurred as a result of lockdown, such as inflation, damage to the economy and the tourism sector. We see three different types of discussions for low income, middle income and high income countries respectively: 1) lockdown effects on labour, remittances, oil prices, 2) job market crisis, fragile economy and lack of trade, and 3) quarantine issues, protests and hot-spots lockdown issues.

**Seriousness of COVID-19 virus:** Terms that appeared in newspapers from high income countries show that these countries declare COVID-19 as a Global pandemic. Newspapers from middle income countries report that COVID-19 has reduced business activities (less travellers), which has resulted in unemployment. They also report shortages of medical staff as a result of contact with COVID patients. On the other hand, newspapers from low income countries report that COVID-19 has severally effected the remittances, earnings of families which results in difficulty accessing essential goods.

**Can mask protect against COVID?:** Low income countries' newspapers discuss the usage of masks at places like restaurants, bars, and laboratories. They also discuss mask-shortage issues. News from middle income countries present keywords such as entertainment halls, banquet halls, and restaurants. On the other hand, news from high income countries mention the benefits of mask use.

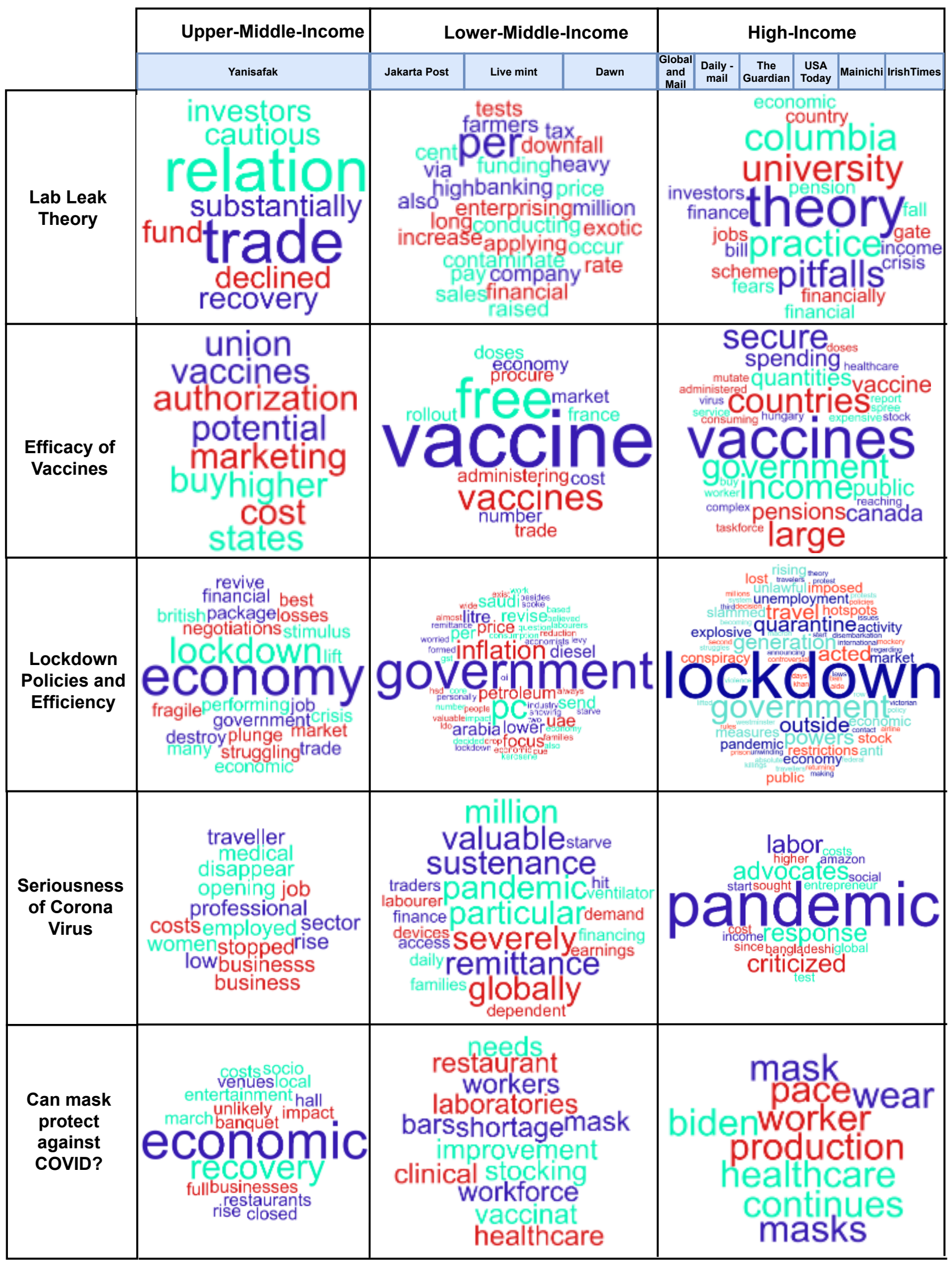

**FIGURE 6. Word clouds showing the keywords appearing most frequently related to five queries in relation to different economies.**

### *C. DISCUSSIONS EVOLVED DURING DIFFERENT STAGES OF COVID-19:*

The results showed that the extracted topics correspond to the chronological development of the pandemic and the measures that, according to our research queries, were under discussion in different countries and across different economies and political alignments. Figure 7 shows the trends of different economic and political topics. We have identified nine different groups of topics that in our data had similar trends over time: 1) Restrictions (lockdown), oxygen, ventilators, Allergies, Reusable masks, 2) Protests, rally, 3) Pneumonia, dose, rollout, 4) Effects, origin, emerge, lab, misinformation, breath, and identify, 5) Unemployment, scheme, labor, bars, restaurants, protests, 6) Trade, investors, price, stock, 7) Farmers, earning, banking, hot spot, 8) Travellers, rollout, and 9) Practice, fund, marketing, hall. By observing the trends of the discussion topics, we see that some topics are present to some extent over time, such as restrictions and ventilators. There are some other topics that where very frequent in news at the beginning of 2020 and lost popularity over time, such as investors and pneumonia, while others gained popularity, such as unemployment and protests. A closer look in Figure 7 shows a few groups of topics whose frequency is dropping over time. For instance, groups 3), 4), 6), 8) which cover topics such as rollout, effects, misinformation, breath, trade, price, travellers. On the other hand, group 2) and 7) with topics such as protests, rally, earnings, banking have increasing popularity over the observed time period.

**TABLE 1. This table shows the predominant sentiments (Negative, Positive, Fluctuation) for each newspaper during different quarters (2020 - 2021).**

| Newspapers | 1/2020 - 4/2020 | 5/2020 - 8/2020 | 9/2020 - 12/2020 | 1/2021 - 5/2021 |
|---|---|---|---|---|
| Dawn.com | Negative | Negative | Positive | Fluctuation |
| Thejakartapost.com | Negative | Fluctuation | Positive | Fluctuation |
| Livemint.com | Negative | Positive | Positive | Positive |
| Mainichi.jp | Negative | Fluctuation | Positive | Fluctuation |
| Yenisafak.com | Fluctuation | Negative | Fluctuation | Fluctuation |
| Theglobeandmail.com | Negative | Fluctuation | Positive | Positive |
| USAToday.com | Negative | Positive | Positive | Positive |
| Dailymail.co.uk | Negative | Positive | Positive | Positive |
| Theguardian.com | Negative | Positive | Positive | Positive |
| Irishtimes.com | Negative | Positive | Positive | Positive |

### *D. SENTIMENT SCORE DURING DIFFERENT STAGES OF COVID-19:*

The data on COVID-19 as reported by the European Centre for Disease Prevention and Control shows that Europe is affected more than Asia and North America (respectively) in terms of total positive cases and deaths per million and total tests per thousand. Overall, from January 2020 to April 2020, the pattern of sentiment in news about the outbreak of COVID-19 disease was the same in Asia, Europe and North America (see Figures 8, 10, 9). However, from May 2020 to July 2021, sentiment score was low in Europe and North America whereas in Asia it varied on the positive side. Onward until December 2020, sentiment score was negative for Europe only. From January 2021 to May 2021, the sentiment score was negative in Europe and Asia whereas it remained positive in North America.

Figure 8 shows the comparison of sentiments across Asia, utilizing the news articles. Each line graph shows the SA of news reported by top five newspapers (Dawn, The Jakarttapost, Livemint, Mainichi, Yanisafak), covering a period from the January 2020 to May 2021. In Asia, the sentiment score increased from negative to positive in articles published by Dawn, The Jakarttapost, Livemint, Mainichi newspapers, while in the case of Yanisafak it stayed negative or fluctuated. In the first quarter (January 2020 - April 2020), the reported sentiment was mostly negative for all the newspapers except Yanisafak which fluctuated between positive and negative (see Table 1). In the second quarter (May 2020 - August 2020), the reported sentiment was approximately positive for Livemint, negative for Dawn and Yanisafak and there were fluctuations for almost all the newspapers. In the third quarter (September 2020 - December 2020), the reported sentiment was approximately positive for all the newspapers except Yanisafak which fluctuated between positive and negative. In the first quarter of 2021 (January 2021 - May 2021), there were fluctuations for mostly all the newspapers.

In North America, the sentiment score began with negative scores and then changed to positive score dramatically for news reported by all the newspapers in North America (Theglobemail and USAToday). We located the reasons for this change by looking the most frequent words (see Table 3). In fact, in the beginning, there were more negative words such as outbreak, killed, risk, and coronavirus whereas after 4 months there were less negative words such as virus, and pandemic. At the start of the period (January 2020 - April 2020), the reported sentiment was mostly negative for both newspapers. In the second quarter (May 2020 - August 2020), the reported sentiment was approximately positive for USAToday and fluctuated between negative and positive for Theglobemail. In the third quarter of 2020 (September 2020 - December 2020) and in the first quarter of 2021 (January 2021 - May 2021), the reported sentiment was approximately positive for both the newspapers.

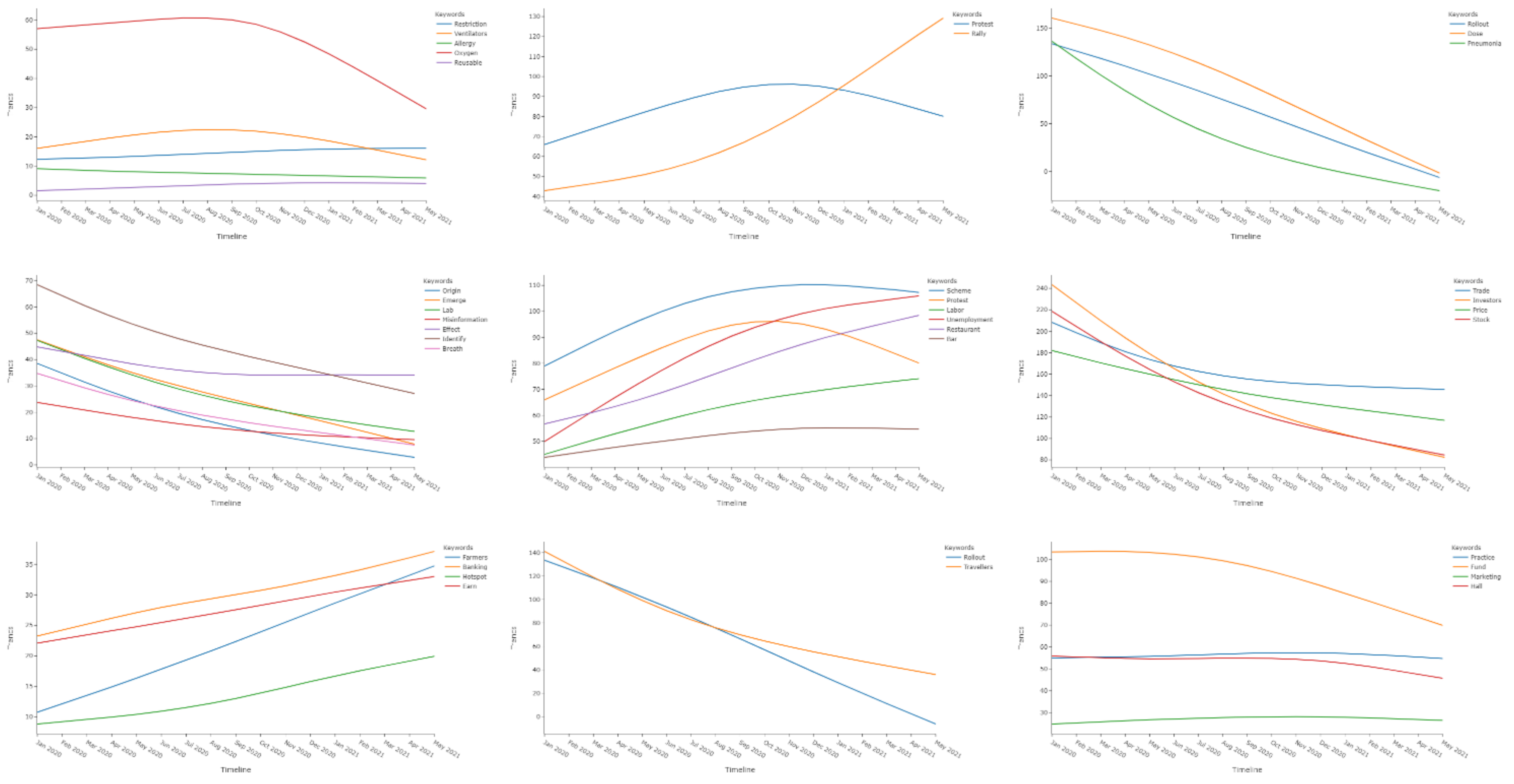

**FIGURE 7. Graphs showing nine groups of topics with similar trends (x-axis = time, y-axis = trends). From top left 1) Restrictions, next to it 2) Protests and 3) Pneumonia, in the second row 4) Effects, 5) Unemployment, 6) Trade, in the bottom row 7) Farmers, 8) Travellers, 9) Practice.**

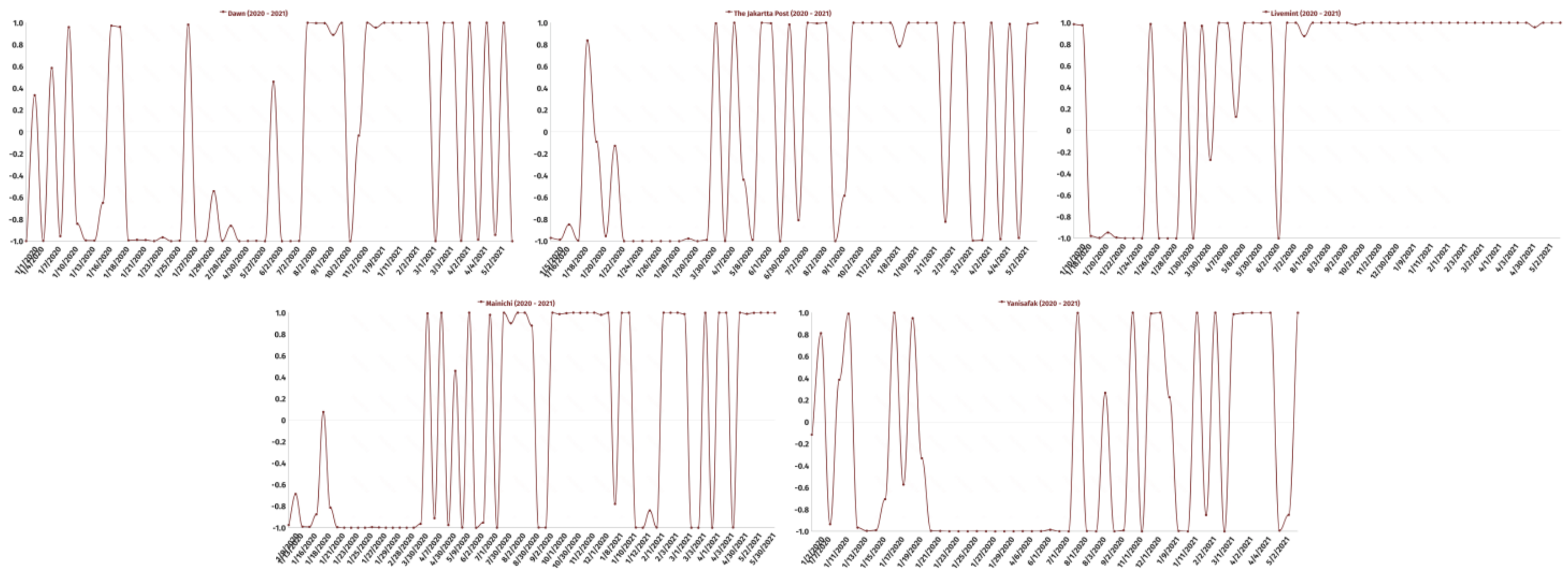

**FIGURE 8. Comparison of sentiment score on scale -1 to 1 over months in news reported by top newspapers in Asia related to COVID-19. Each dot represents the sentiment of news articles published on specific date. The shown sentiment score varies: first quarter (January-April 2020) mostly negative, second quarter (May-August 2020) mostly positive, third quarter (Sep-Dec 2020) mostly positive and, fourth and last quarter (Jan-May 2021) fluctuation between positive and negative.**

On the other hand, the sentiment score of news reported by European newspapers (Dailymail, The Guardian, Irishtimes) was negative in the first quarter of 2020 (January 2020 - April 2020), whereas in all other 3 quarters (May 2020 - August 2020, September 2020 - December 2020, January 2021 - May 2021), the score mostly stayed positively.

## V. DISCUSSION

Overall, the results of our spatio-temporal analysis confirm with our research hypothesis. The general findings on how political and economic issues propagated over time during the pandemic and across different political and economic barriers provided relevant insights. Left wing newspapers promote a degree of social equality (see Table 2). What we found as common and most frequent topics were emergence of virus, its causes, questions about efficacy of vaccines, effects of lockdown, financial aspects of the market, and reusable masks. Right wing newspapers support free markets and reduction of government spending (see Table 2) and the list of topics that appeared most frequently were conspiracy and misinformation about Coronavirus, protests, policies on face masks. The newspapers with moderate political alignment do not stand on either side of the spectrum (left or right wing) according to the Table 2. Since the topic COVID-19 has different associated issues it's very hard to describe what is a moderate topic during the pandemic. But overall we see topics such as conspiracy and misinformation about virus, symptoms of the virus such as breathing, heart problem, pneumonia, respiratory system diseases, allergies and need for oxygen. Generally, liberal newspapers support more governmental involvement, and subsidized public facilities such as public education (see Table 2). What we saw as most prominent was forecasting about vaccine roll-out, different vaccines with different doses and for different age groups, and the importance of ventilators. The overall findings on economic issues which appear as a consequence of COVID-19 are quite consistent with the economic situation of the country of the newspaper. Additionally those newspapers that focus on local political content rather global do the same when it comes to the economy. In newspapers from lower level economies the most frequent topics were funding, pays, pricing, farmers, banking, vaccine cost, marketing, lockdown effects on labour, remittance, oil prices, and starvation. In case of middle level economies the most frequent topics were funds, trade, investment, job crisis, no trade, lower business, unemployment and fragile economy. For high income economies the most frequent topics were scheme, vaccines, number of doses, stocking the vaccine, task forces, quarantine issue, protests, global pandemic, and hot-spot lockdown issues.

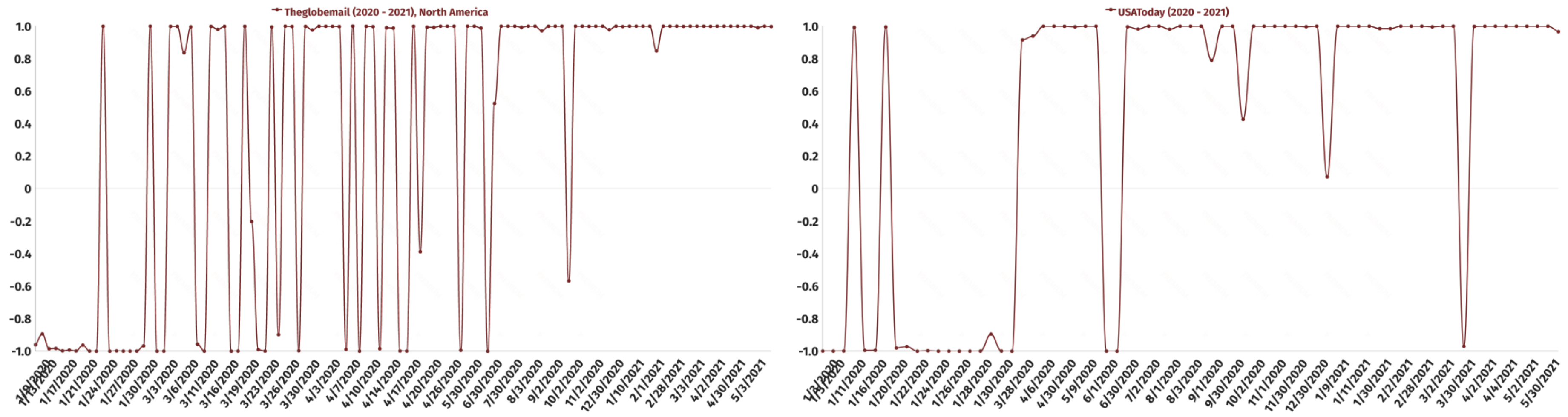

**FIGURE 9.** Comparison of sentiment score on scale -1 to 1 over months in news reported by top newspapers in North America related to COVID-19. Each dot represents the sentiment of news articles published on specific date. The shown sentiment score varies: first quarter (January-April 2020) negative, second quarter (May-August 2020) mostly mostly positive, third quarter (Sep-Dec 2020) positive and, fourth and last quarter (Jan-May 2021) positive.

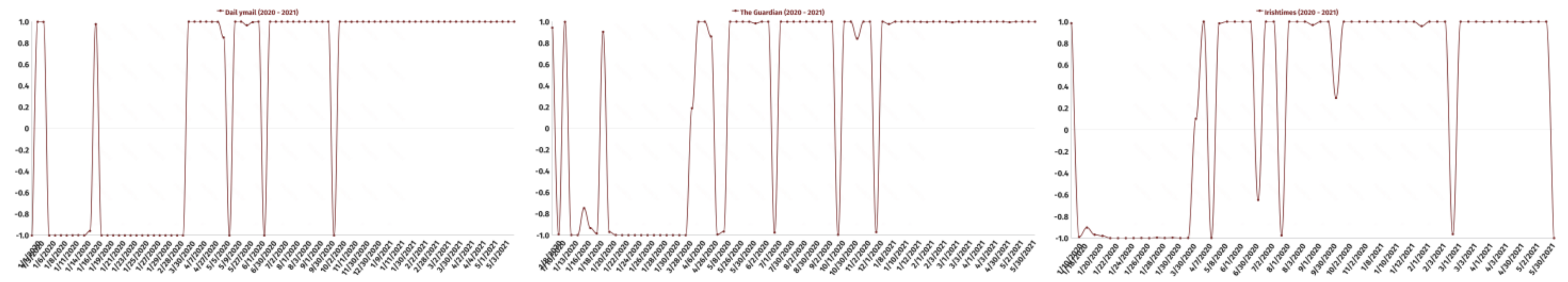

**FIGURE 10.** Comparison of sentiment score on scale -1 to 1 over months in news reported by top newspapers in Europe related to COVID-19. Each dot represents the sentiment of news articles published on specific date. The shown sentiment score varies: first quarter (January-April 2020) negative, second quarter (May-August 2020) mostly positive, third quarter (Sep-Dec 2020) positive and, fourth and last quarter (Jan-May 2021) positive.

Regarding the second research question - How different discussions evolved during different stages of the COVID-19 epidemic? -, we see that the trends of different correlated terms were same although the frequency of occurrence of the topics were different. For example, origin, emerge, lab, misinformation, and so one has had same trend over time but their frequency of occurrence were different (see Figure 7).

Regarding the third research question - What are the patterns of emotional states during different stages of the pandemic across different countries? -, we see that the news published by European and North American newspapers depict more or less the same sentiments, whereas the Asian newspapers show some differences.

## VI. CONCLUSIONS AND FUTURE WORK

In this paper we focused on the analysis of news articles related to COVID-19 (published between January 2020 and May 2021) by observing the political alignment of different newspapers, different economies, and sentiments across different continents. Our prime motivation was to understand news reporting by newspapers with different political alignments and news from different economies. We also include the analysis of topics which have had similar trends over time and comparison of sentiments across different continents. We firstly constructed five queries related to COVID-19 and filtered news articles for these queries for each month (total 17 months) and for each newspaper. We performed TM and generated topics. Afterward, we filtered political and economic issues and analyzed the topics on top of political alignments and different economies (see Figure 6 and 5). Next, we identified the trends of similar terms (see Figure 7). Lastly, we calculated sentiment scores for each newspaper using each news article (see Figures 8, 10, 9). Our findings suggest that 1) Left wing newspapers report on raising questions and are futured-oriented, 2) The newspapers with moderate political alignment report on conspiracies and misinformation, 3) Right wing newspapers report on conspiracies and misinformation as well as protests and rallies, 4) Liberal newspapers report on worldwide challenges such as lockdown and face masks, 5) Regarding the economy, we see that newspapers report on national economic situations more than on global economic situations, 6) With regards to sentiment, we see that the news published by European and North American newspapers depict more or less the same sentiments, whereas the Asian newspapers show some differences.

Overall, our study suggests that the political alignment of a newspaper and the economic condition of a country influence news spreading. For example, topics such as lockdown effects on labour, vaccine cost, prices appear at the same time in newspapers from lower level economies, whereas topics such as number of doses, task forces, quarantine protests appear at the same time in newspapers from high income economies. If we look at the evolution of topics we see that topics which are semantically related to each other have similar trends over time. For instance, topics such as origin of virus, lab theory, misinformation appear at the same time in newspapers. Similarly other topics such as unemployment, labor, protests appear at the same time in newspapers. Our analysis of sentiment score suggests that the news articles published by European and North American newspapers depict more or less the same sentiments whereas Asian newspapers show greater differences. We verified that sentiment score in European newspapers was negative in the beginning of the pandemic, but became positive in the rest of the pandemic. In the case of the Asian newspapers the differences between newspapers are more significant, and it is more difficult to devise a line of concordance in sentiment amongst newspapers, particularly over time.

There are a few limitations that we compromised on while analyzing the results and answering our research questions. Data size is not big enough. The preconditions to select the newspaper (see Section III-A) are very strict because the platform that we use to collect news articles (Event Registry) does not provide enough news articles for all the newspapers or it might be possible that some newspapers do not make their article available after some time period or do not publish enough articles related to COVID-19. Another reason is that we estimate the political alignment using Wikipedia-infobox which does not provide political alignment for all the newspapers. Also, there are multiple determinants to know the economic conditions of a country [5], [6], and we only use income level to estimate the economic context of a country.

In conclusion, news offers detailed information about the current pandemic situation in different countries. Insights gained from this analysis can support government decision making and communication strategies. It can also encourage further discussion about the management of COVID-19 and other global health events in accordance to each country's political and economical context. In the future, we plan to analyze news reporting differences on top of different geographical places, different cultural values, and linguistic influences.

**TABLE 2. Political alignments (PA) of top newspapers along with definition of each PA. As RQ1 (see SectionI-C) finds political issues across different political contexts therefore this table reminds of political alignment of each newspaper and its definition.**

| Political Alignment | Newspaper | Description |
|---|---|---|
| **Liberal** | **Dawn** | • More governmental involvement to promote socio-economic equality.<br>• Gradual speed of change in government.<br>• The government subsidizes public education at the college level. |
| **Socially Liberal** | **The Irish Times, Live Mint** | • Expected to address economic and social issues such as poverty, welfare, infrastructure, health care.<br>• Emphasize the rights and autonomy of the individual. |
| **Centre Left** | **Mainichi Shimbun, The Guardian, The Jakarta Post** | • It promotes a degree of social equality.<br>• It opposes a wide gap between the rich and the poor.<br>• Support moderate measures to reduce the economic gap such as progressive income tax |
| **Islamic Conservative** | **Yeni Şafak** | • It applies the teachings of particular religions to politics.<br>• It oppose abortion, drug use, and sexual outside of marriage. |
| **Centre Right** | **Daily Mail, The Globe and Mail** | • It supports free markets, limited government spending. |
| **Centrist** | **Dawn, USA Today** | • It leans to accept or support a balance of social equality and a degree of social hierarchy.<br>• It opposes the political changes that can shift society either to the left or right. |
| **Left Wing** | **Mainichi Shimbun, The Guardian, The Jakarta Post** | • It supports social equality (in a society, equal rights, liberties and status, freedom of speech, etc.) |
| **Right Wing** | **Daily Mail, The Globe and Mail** | • It supports the view that certain social orders and hierarchies are inevitable, natural, normal, or desirable. |

## VII. ACKNOWLEDGMENTS

The research described in this paper was supported by the Slovenian research agency under the project J2-1736 Causalify and by the European Union's Horizon 2020 research and innovation program under the Marie Skłodowska-Curie grant agreement No 812997.

[5] https://www.prosperity.com/rankings

[6] https://datahelpdesk.worldbank.org/knowledgebase/articles/906519-world-bank-country-and-lending-groups

**TABLE 3. Most frequent words in different time periods representing sentiment. The sentiment is based on body text of a news article and we show the most frequent words during different time.**

| | | | | |
|---|---|---|---|---|
| **Dawn.com** | virus, cases, outbreak, lockdown, infected, disease, | patients, death | prime, economic, national time, pandemic, positive, issue,pilots, vaccine, support, debate, political, gas, help | spread, authorities, outbreak, health, city, measures, travel, situation, case, global, virus, market, passengers, warn~ |
| **Thejakartapost.com** | outbreak, spread, infected, killed, virus, disease, virus, tourists, coronavirus, medical, flights, respiratory | city, infections, home, authorities, ~restrictions, measures, president | pandemic, economic, year, social, vaccine, national, data, local, help, positive, tested | Wuhan, outbreak, spread, infected, citizens, virus, disease, deadly, ~suspected, world, virus, tourists, market |
| **Livemint.com** | china, outbreak, spread, infected, confirmed, markets, flights, disease, prices, death, situation, workers, investors, deadly | demand, rate, total, growth, take, country, social, testing, patients, positive, virus, vaccine, business, recovery, tested, ~increase, high,~ | vaccine, growth, pandemic, data, demand, sales, positive, recovery, ~higher, increase, rate, economy, business,~ | china, virus, outbreak, spread, travel, ~impact, flights, measures, novel, medical, patients, situation, workers, investors, international |
| **Mainichi.jp** | outbreak, authorities, travel, medical, measures, virus, patients, respiratory, symptoms, severe, emergency, companies, deadly | social, police, novel, masks, Chinese, week, health, testing, police, business, national | Health, spread, outbreak, travel, medical, measures, flights, infections, ~president, voters, economy, positive, vote | China, Wuhan, spread, outbreak, china, infected, travel, medical, measures, global, respiratory, symptoms, emergency, cause, deadly |
| **Yenisafak.com** | outbreak, India, flights, internet, infected, travel, virus, Pakistan, virus, news, Respiratory, ~Korea, media, Organization | spread, least, recoveries, pandemic, nuclear, killed, coronavirus, hit | total, pandemic, past, positive, ~infections, vaccine, economic, tested, deaths, recovery, restrictions | outbreak, spread, flights, coronavirus, respiratory, international, medical, lockdown, ~statement, world, capital |
| **Theglobeandmail.com** | china, global, spread, outbreak, ~health, markets, prices, stock, price, ~reported, risk, | police, workers, virus, long, week, home, long | global, market, election, vaccine, across, investors, pandemic, financial,~ | virus, global, market, spread, outbreak, health, financial, prices, impact, stock, price, risk |
| **USAToday.com** | china, travel, spread, outbreak, ~health, passengers, news, flights, respiratory, symptoms, severe, killed, coronavirus | social, police, players, virus, pandemic, right, care, school, work, week | positive, election, white, tested, pandemic, early, made, season, test, mask, election, mask, way, country | Virus, spread, travel, medical, flights, impeachment, world, city, Wuhan, Chinese, passengers |
| **Dailymail.co.uk** | Wuhan, outbreak, patients, infected, ~symptoms, medical, tested, deadly, disease, virus, coronavirus,~ | Chinese, outbreak, city, health, symptoms, medical, travel, masks, tested, man, virus, world, postponed, positive | lockdown, make, set, work, pandemic, restrictions, see, positive, social, support, government | spread, patients, health, travel, tested, disease, world, authorities, postponed, positive |
| **Theguardian.com** | virus, outbreak, spread, medical, ~travel, disease, symptoms, infected, risk, death, infected, risk, China | social, pandemic, think, local, support, really, police, working, restrictions | support, national, care, election, ~restrictions, social, economic, England | Medical, Health, Travel, British, City, Global, Staff, country, death, family, case, risk, citizens |
| **Irishtimes.com** | China, spread, outbreak, city, patients, global, infected, death, emergency, market, fell, medical, measures, hospital, WHO | Work, Back, Home, Family, Go, Social, Travel, Children, Country, restrictions, pandemic, life, lockdown | Work, Group, Good, Well, Positive, ~Months, Chief, Government, Business, Tested, European | Wuhan, Spread, travel, outbreak, health, city, patients, global, infected, staff, death, emergency, market, fell, measures, citizens, deaths |

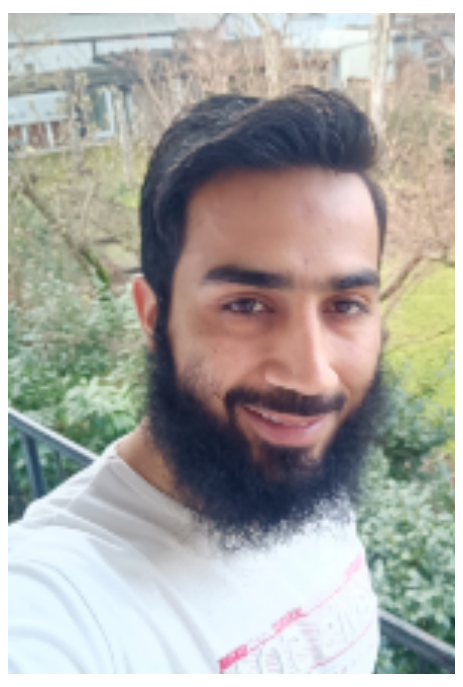

ABDUL SITTAR is pursuing Ph.D. at Jozef Stefan Institute, Slovenia and also a MSCA fellow currently working on CLEOPATRA ITN. He acquired MS degree from COMSATS Institute of Information Technology, Lahore, Pakistan. His research work focus on Natural Language Processing (NLP) and machine learning (ML). He passed BS in computer science from University of Gujrat, Gujrat, Pakistan. From 2013 to 2017, he was working as a software engineer in Nextbridge (Pvt.) Ltd. His development career includes iOS and Android mobile application development. From 2018 to 2019, he was working as lecturer at PMAS-Arid Agriculture University Rawalpindi.

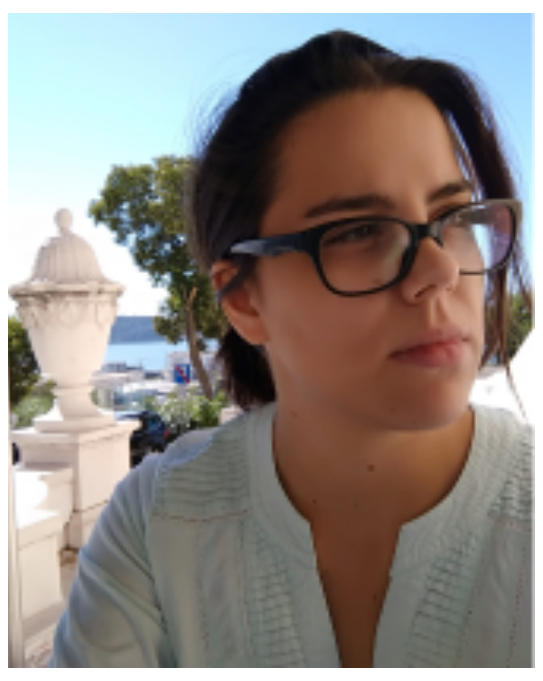

DANIELA MAJOR received a B.A degree in History, a M.Litt in Intellectual History, and is currently pursuing a PhD in Digital Humanities at School of Advanced Study, University of London. From 2018 to 2019, Ms. Major was a research scholar at Arquivo.pt, the Portuguese Web Archive, where she used multilingual sources kept in web archives to study the Commemorations of the First World War. Currently, she is working on a thesis on Ideas of Europe and the modern media coverage of the European Union. Her interests include contemporary European History, conceptualizing and working with born-digital sources, and the preservation of online content.

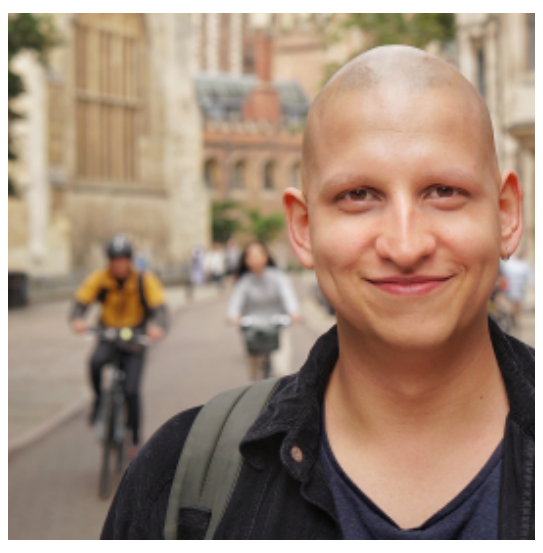

CAIO MELLO is a Ph.D. student in Digital Humanities at the School of Advanced Study, University of London. He holds a BA in Journalism and a MA in Communication, both from the Universidade Federal de Pernambuco, Brazil. He works as an early-stage researcher at the University of London for the EU-funded Horizon 2020 project CLEOPATRA, under the Marie Skłodowska-Curie Innovative Training Network. Before starting his PhD, he was a fellow researcher at CAIS - Center for Advanced Internet Studies in Bochum, Germany.

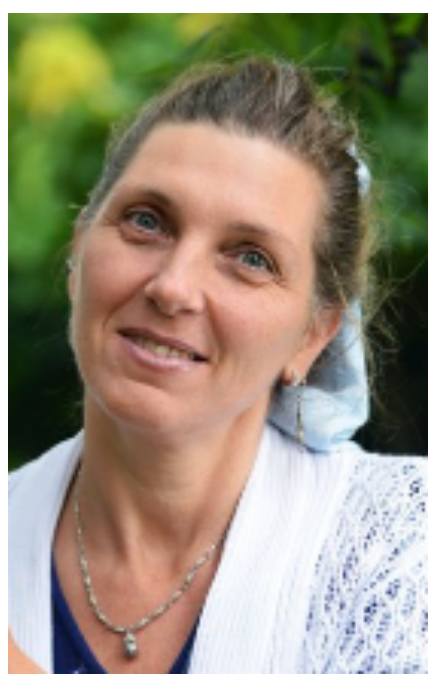

DUNIA MLADENIĆ is leading Artificial Intelligence Laboratory at Jozef Stefan Institute, serving on the Institute's Scientific Council (2013-2017) as a vice president (2015-2017) and working on a number of research projects mainly related to Machine Learning, Data and Text Mining, Big Data Analytic, Semantic Technologies and their application on real-world problems. Professor at Jozef Stefan International Postgraduate school, and University of Ljubljana, University of Primorska, University of Zagreb (FERI and FOI), teaching classes related to Data Analytics and Artificial Intelligence.

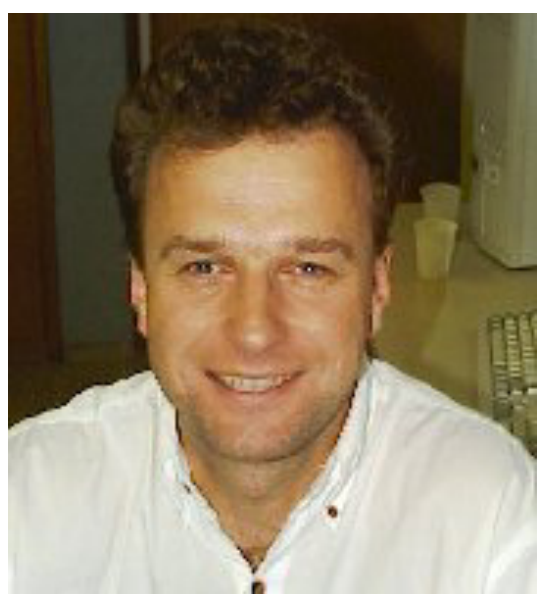

MARKO GROBELNIK is an expert researcher in the field of Artificial Intelligence (AI). Focused areas of expertise are Machine Learning, Data/Text/Web Mining, Network Analysis, Semantic Technologies, Deep Text Understanding, and Data Visualization. Marko co-leads the Department for Artificial Intelligence at Jozef Stefan Institute, co-founded UNESCO International Research Center on AI (IRCAI), and is the CEO of Quintelligence.com specialized in solving complex AI tasks for the commercial world. He collaborates with major European academic institutions and major industries such as Bloomberg, British Telecom, European Commission, Microsoft Research, New York Times. Marko is co-author of several books, co-founder of several start-ups and is/was involved into over 50 EU funded research projects in various fields of Artificial Intelligence.

• • •